\documentclass[aps,prl, twocolumn, groupedaddress,showpacs]{revtex4-1}
\usepackage{graphics}
\usepackage{epsfig}

\begin{document}

\title{Elastic and structural instability of cubic $\mathrm{Sn_3N_4}$ and $\mathrm{C_3N_4}$ under pressure}

\author{Gopal K. Pradhan,$^1$ Anil Kumar,$^2$ Umesh V. Waghmare$^2$ and Chandrabhas Narayana$^1$}l
\email[Address for correspondence ]{cbhas@jncasr.ac.in}
\affiliation{$^1$Light Scattering Laboratory, Chemistry $\&$ Physics of Materials Unit, $^2$Theoretical Sciences Unit, Jawaharlal Nehru Centre for Advanced Scientific Research (JNCASR), Bangalore 560064, India}
\author{Sudip K. Deb}
\affiliation{Indus Synchrotron Utilization Division, Centre for Advanced Technology, Indore 452013, India}

\begin{abstract}
We use in-situ high pressure angle dispersive x-ray diffraction measurements to determine the equation of state of cubic tin nitride ($\gamma$-$\mathrm{Sn_3N_4}$) under pressure up to about 26 GPa. While we find no evidence for any structural phase transition, our estimate of the bulk modulus ($B_0$) is 145 ($\pm$ 1.7) GPa, much lower than the earlier theoretical estimates and that of other group IV-nitrides.  We corroborate and understand these results with complementary first-principles analysis of structural, elastic and vibrational properties of group IV-nitrides, and predict a structural transition of $\mathrm{Sn_3N_4}$ at a higher pressure of 88 GPa compared to earlier predictions of 40 GPa. Our comparative analysis of cubic nitrides shows that bulk modulus of cubic $\mathrm{C_3N_4}$ is the highest (379 GPa) while it is structurally unstable and should not exist at ambient conditions.
\end{abstract}
\maketitle

Since the discovery of cubic spinel ($\gamma$) phases in Group IV nitrides\cite{tin,ger,sil}, they have received a renewed interest in the past few years due to prediction of superhardness\cite{cohen} in these class of materials. It is also expected that these spinel classes of nitrides will exhibit interesting electronic properties, such as varying electronic band gap energies suitable for optoelectronic applications. So far, the binary nitrides of tin\cite{tin}, germanium\cite{ger} and silicon\cite{sil} have been synthesized and shown to adopt the spinel structure under different experimental conditions. These new polymorphs have the dense spinel structure with Si/Ge/Sn atoms in octahedral as well as tetrahedral coordination with nitrogen and the coordination of nitrogen is four. The crystal structure of $\gamma$-$A_3\mathrm{N_4}$ ($\gamma$-$A_3\mathrm{N_4}$, $A$ = Si, Ge and Sn) (space group $Fd\bar{3}m$, \#227) can be described as a distorted close-packed lattice of nitrogen atoms at (${u, u, u}$) positions with $A$ atoms occupying 1/8 of the tetrahedral interstitial sites and 1/2 of the octahedral sites\cite{tin,ger,sil}. The anion arrangement produces a rigid-vertex linkage of $A$-filled regular tetrahedra ($A^t\mathrm{N_4}$) and distorted octahedral ($A^\mathrm{o}\mathrm{N_6}$) sharing half of their edges. The cation sublattice can be described as an array of empty corner-sharing tetrahedra and $A$-filled truncated tetrahedral. The bulk modulus of the spinel phase is expected to be high, primarily determined by the nitrogen sublattice, but it should be influenced by the choice of cations. A very high bulk modulus of 369 GPa\cite{carbon} has been predicted for the cubic phase of $\mathrm{C_3N_4}$, without any success in synthesizing this material. Though both $\mathrm{Si_3N_4}$ and $\mathrm{Ge_3N_4}$ have similar bulk moduli, the same for $\mathrm{Sn_3N_4}$ has been predicted to be significantly different. Also, from recent nanoindentation studies\cite{nano}, tin nitride spinel was found to be significantly softer and more plastic than its lighter congeners ($\mathrm{Si_3N_4}$ and $\mathrm{Ge_3N_4}$). Thus, the exact role of the cation in determining the stability and elastic properties is not yet completely understood.

Tin nitrides are of special interest due to their promising semiconducting and electrochromic properties\cite{tnsem,tnel}. The feasibility of utilizing tin nitride thin film as write-once, optical recording media has also been reported\cite{tnor}. Therefore, the precise knowledge of the stability and mechanical properties of $\gamma$-$\mathrm{Sn_3N_4}$ would be of great interest for several practical applications. Though the ambient bulk modulus ($B_0$) of $\gamma$-$\mathrm{Sn_3N_4}$ and $\gamma$-$\mathrm{Ge_3N_4}$ has been intensively investigated\cite{si1b,si2b,geb}, there have been no experimental studies to determine the compressibility of tin nitride. With lack of experimental data, first-principles calculations have been used to investigate the properties of $\mathrm{Sn_3N_4}$\cite{nano,tn1,tn2,tn3,tn4}. However, the predicted values of the bulk modulus appear to vary between 186 and 218 GPa. In this work, we have performed high pressure x-ray diffraction measurements on $\gamma$-$\mathrm{Sn_3N_4}$ to determine the bulk modulus from the volume as a function of compression. We have also carried out first-principles density functional theory (DFT) calculations in the generalised gradient approximation (GGA) to achieve a consistent understanding of our experimental findings. First-principles calculations have been carried out for other group-IV nitrides as well to (a) determine the stability of the spinel phase with respect to pressure and possible phase transitions, and (b) understand the influence of cation on elastic behaviour. Based on our calculations, we also discuss the stability of cubic phase of $\mathrm{C_3N_4}$ at ambient conditions.

Spinel tin nitride ($\mathrm{Sn_3N_4}$) was synthesised by high-pressure solid-state metathesis reactions where tin tetraiodide was reacted with lithium nitride and ammonium chloride in a piston-cylinder apparatus at 623 K and 2.5 GPa. Details of the synthesis procedure can be found elsewhere\cite{tnsynt}. In-situ high-pressure  angle dispersive x-ray powder diffraction measurements (up to ~26 GPa) were performed using the monochromatic synchrotron radiation ($\lambda$ = 0.68881 {\AA}) at XRD1 beamline at ELETTRA synchrotron source, Italy. The two-dimensional x-ray diffraction (XRD) patterns were collected on a MAR345 imaging plate. The sample-to-detector (image plate) distance was calibrated by collecting the diffraction pattern of powdered silicon (Si) at ambient conditions. The one-dimensional diffraction patterns were obtained by integrating along the Debye-Scherrer ring in the two dimensional image patterns using the FIT2D software\cite{FIT2D}. Polycrystalline $\mathrm{Sn_3N_4}$ powder was placed together with gold into the 200 $\mu$m hole of a stainless steel gasket (T301), preindented to 60 $\mu$m, inserted between the diamonds of a Mao-Bell-type diamond anvil cell (DAC). Gold was used as an internal pressure calibrant and the pressure in the DAC was determined using the known equation of state of gold \cite{gold}. Methanol-ethanol-water (MEW) mixture (16:3:1) was used as pressure transmitting medium. To develop a microscopic understanding of the discrepancy between the calculated bulk modulus earlier\cite{nano,tn1,tn2,tn3,tn4} and our experimental estimate, we present new first-principles calculations based on DFT with a generalized gradient approximation (GGA)\cite{gga} for the exchange correlation energy, as implemented in PWSCF\cite{PW} package. We use ultrasoft pseudopotentials\cite{soft} to describe the interaction between the ionic core and valence electrons, and a plane wave basis with energy cutoffs of 30 Ry and 240 Ry in representation of wave functions and charge density respectively. Integrations over Brillouin zone are sampled with an $8\times8\times8$ Monkhorst Pack Mesh\cite{mesh}. Structural optimization is carried through minimization of energy using Hellman-Feynman forces in the Broyden-Flecher-Goldfarb-Shanno (BFGS) based method. Complete phonon dispersion is determined using DFT-linear response and $2\times2\times2$ mesh-based Fourier interpolation of interatomic force constants.

\begin{figure}[t]
  \includegraphics[scale=0.30]{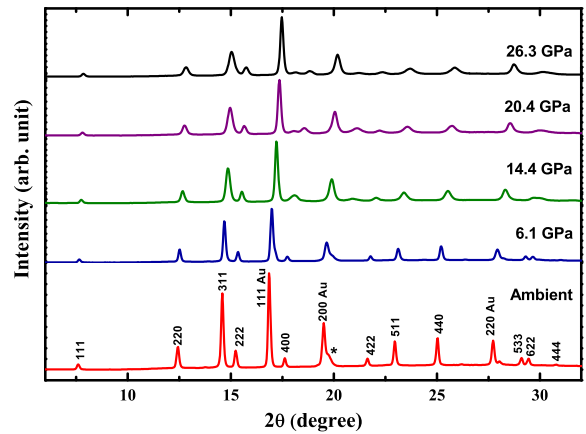}\\
  \caption{Pressure evolution of X-ray diffraction patterns of $\gamma$-$\mathrm{Sn_3N_4}$ compressed quasi-
hydrostatically in MEW up to 26 GPa. Diffraction lines due to gold (Au) (used for in-
situ pressure calibration) and stainless steel gasket (*) are also identified.}\label{*}
\end{figure}

Figure 1 shows the XRD patterns of $\gamma$-$\mathrm{Sn_3N_4}$ at various pressures. The cell parameters were obtained by analysing the XRD profiles by Le-Bail profile fit using the GSAS software\cite{GSAS}. Ambient lattice parameter has been found to be $a$ = 9.0205(5) {\AA}, which is in close agreement to earlier reported value\cite{tnsynt}. Upon increasing the pressure, we didn't observe any new diffraction peaks (see Fig. 1) or sudden discontinuity in the pressure dependency of the $d$ values up to the maximum achieved pressure of ~26 GPa. This indicates that $\gamma$-$\mathrm{Sn_3N_4}$ doesn't undergo any transition and remains in the cubic crystal structure up to the highest pressure. Figure 2 shows pressure volume compression data. Using the second order Birch-Murnaghan ($B_0^{'}$ = 4) equation of state (EOS)\cite{BM} to fit the volume compression data, we found the bulk modulus ($B_0$) to be 145 ($\pm$1.7) GPa. It is interesting to note that the value thus obtained is much lower than the predicted bulk modulus values. Thus it was necessary to revisit the theoretical calculations.

\begin{figure}[t]
\includegraphics[scale=0.30]{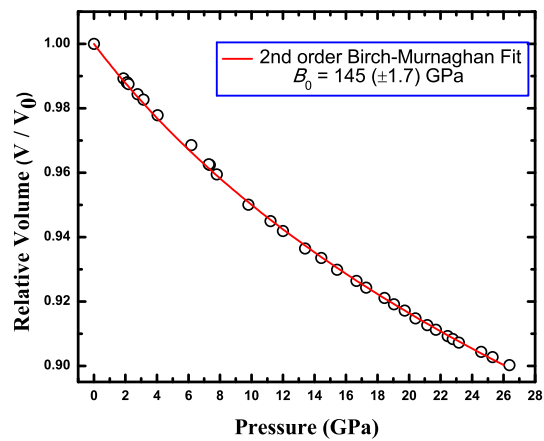}
\caption{Relative volume change in $\gamma$-$\mathrm{Sn_3N_4}$ as a function of pressure at ambient temperature (Solid line is a second order Birch-Murnaghan fit to the experimental data (open symbols)).}
\end{figure}

Our theoretical estimate of the lattice constant of $\gamma$-$\mathrm{Sn_3N_4}$ is 9.136 {\AA}, within typical GGA errors of the experimental value of 9.0205 {\AA}\cite{tnsynt}. Bulk modulus estimated from first-principles energies as a function of volume is 158 GPa, in reasonable agreement with our measured value of 145 GPa reported here. We note that the earlier theoretical estimate\cite{tn4} of the bulk modulus is within local density approximation (LDA) (contrary to the claim of use of GGA in this paper; we have reproduced their result with the same code and LDA choice of energy functional), and is overestimated to be 187 GPa. Thus, the observed softness of $\mathrm{Sn_3N_4}$ reflected in the lower value of \emph{B} here should be reliable. With the same choice of exchange correlation energy functional, we systematically studied properties of other group IV nitrides in the spinel structure, and find that the bulk modulus (see Table I) reduces from 379 GPa of carbon nitride to 158 GPa of $\mathrm{Sn_3N_4}$. While the lattice constant increases from carbon to tin by almost 30\%, consistent with the size of the cation, the internal structural parameter ($u$) changes by only 2$\%$.

\begin{table}[h]
\caption{Experimental and calculated equilibrium lattice constant ($a_0$) and bulk modulus ($B$) of the $\gamma$-phase of different group-IV nitrides.\label{}}
\begin{tabular}{|c c c c c c|}
\hline
& Calc.$a_0$ & Expt.$a_0$ & Calc.$B$ & Expt.$B$ & $u$ \\
& ({\AA}) & ({\AA}) & (GPa) & (GPa) &  \\
\hline
$\gamma$-$\mathrm{Sn_3N_4}$ & 9.136 & 9.0144 & 158 & 145 ($\pm$ 1.7) & 0.2592 \\
$\gamma$-$\mathrm{Si_3N_4}$ & 7.792 & 7.7381(2) & 292 & 308($\pm$ 5) & 0.2575 \\
$\gamma$-$\mathrm{C_3N_4}$ & 6.796 & Not synthesized & 379 &  & 0.2559 \\
\hline
\end{tabular}
\end{table}

The local stability of a structure can be confirmed through determination of the full phonon dispersion, and by verifying that there are no unstable modes in the structure. Our results (see Fig. 3) clearly show that $\mathrm{C_3N_4}$ in the spinel structure is unstable with respect to both shear acoustic and optical modes, and should not form. This is consistent with the fact that there are no experimental reports so far on $\mathrm{C_3N_4}$ in the spinel form. We also find that the frequency range (band-width) of acoustic branches reduces from silicon to tin nitrides, confirming their increasing elastic softness. If $K$ is the spring constant or the stiffness of a bond and $a$ is the lattice constant, a simple analysis shows that the bulk or elastic modulus should scale as ${K/a}$. Although the slopes of the acoustic modes in these nitrides seem different in the phonon dispersion (see Fig. 3), variation of \emph{B} with 1/$a$ is almost linear. This means that the bond-stiffness in the nitrides studies here is quite similar, and it is the volume of the unit cell that is responsible largely for the monotonous decrease in the bulk modulus from carbon to tin nitride. As the spinel structure contains cations with tetrahedral and octahedral coordination, it is not straightforward to describe it with a single type of a bond. Since all anions are symmetry equivalent, our reference to bond stiffness here is reasonable in the sense of average bond stiffness.

\begin{table}[b]
\caption{Calculated Born effective charges of the octahedral, tetrahedral coordinated cations and N atoms and electronic dielectric constant in the $\gamma$-phase of different nitrides.\label{}}
\begin{tabular}{|c c c c c |}
\hline
& Z* of & Z* of  & Z* of N & Dielectric  \\
& tetrahedral cation & octahedral cation &  & constant \\
\hline
$\gamma$-$\mathrm{Sn_3N_4}$ & 3.90 & 4.17 & -3.02 & 9.17  \\
$\gamma$-$\mathrm{Si_3N_4}$ & 3.59 & 3.70 & -2.72 & 5.29  \\
$\gamma$-$\mathrm{C_3N_4}$ & 1.90 & 3.98 & -2.45 & 8.99 \\
\hline
\end{tabular}
\end{table}

\begin{figure}[t]
\includegraphics[scale=0.50]{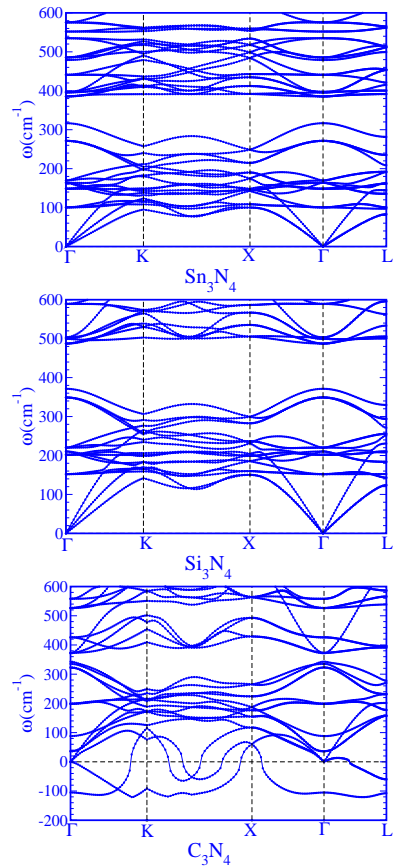}
\caption{Phonon dispersion of  a) $\gamma$-$\mathrm{Sn_3N_4}$, b) $\gamma$-$\mathrm{Si_3N_4}$ and c) $\gamma$-$\mathrm{C_3N_4}$ at equilibrium lattice constant.}
\end{figure}

Our estimates of the Born effective charges also bear a similar feature (see Table II), with two types of effective charges of a cation and degenerate effective charges of N. While the effective charges of $\mathrm{Si_3N_4}$ and $\mathrm{Sn_3N_4}$ are similar, those of $\mathrm{C_3N_4}$ are singularly different. We find that the effective charge of carbon is most anomalous (i.e. deviates most from its nominal charge of Z*=4); this is not very surprising, as the anomalous charges (though larger than the nominal values) are known to correlate with structural instabilities (the presence of unstable vibrational modes) in the system\cite{umesh}. We also note that the effective charges of carbon atoms with 4-fold and 6-fold coordination are rather different, in contrast to relatively similar charges of the two symmetry inequivalent cations in Si and Sn nitrides. We find that the band gap steadily increases from 0.2 eV in $\mathrm{Sn_3N_4}$ to 3.4 eV in $\mathrm{Si_3N_4}$, but reduces substantially to 1.02 eV in $\mathrm{C_3N_4}$, confirming rather different nature of the electronic structure and bonding in $\mathrm{C_3N_4}$ from the rest. Our analysis based on projection of energy eigen-functions on the atomic orbitals shows that carbon in spinel-$\mathrm{C_3N_4}$ is close to being sp$^3$-hybridized with covalent bonding with N, while the cations in $\mathrm{Si_3N_4}$ and $\mathrm{Sn_3N_4}$ appear more ionic. In summary, the unexpectedly low measured bulk modulus of $\mathrm{Sn_3N_4}$ does follow the trend in properties of the group IV-nitrides and seems to be readily understandable.

Our experiments show that $\mathrm{Sn_3N_4}$ remains stable in the spinel structure up to 26 GPa. We now use first-principles simulations to explore its stability at much higher pressures. Since it is difficult to explore all possible structures as a function of pressure, we determine complete phonon dispersion of cubic $\mathrm{Sn_3N_4}$ as a function of pressure. This is an efficient method to find unstable modes and detect any structural transition that the spinel structure would undergo with pressure, and is similar in spirit to approach used commonly in ferroelectric structural transitions\cite{ferro}. Our results (see Fig. 4) show that the $\mathrm{Sn_3N_4}$ remains stable locally in the spinel structure up to pressures of 88 GPa, and develops an instability in the transverse acoustic branch along $<$111$>$ direction. Thus, $\mathrm{Sn_3N_4}$ in the spinel structure will distort with a shear-strain above 88 GPa suggesting a transition from cubic to low symmetry phase. At higher pressures, we find that many more unstable modes develop that involve optical phonons as well, suggesting that the structures beyond 88 GPa are expected to be more complex. Electronic structure of $\mathrm{Sn_3N_4}$ is also rather sensitive to pressure, but exhibits an unusual trend: its band gap increases rather sharply from 0.2 eV at 0 pressure to about 2 eV at a pressure of 50 GPa. This is indeed consistent with the trend in band gaps from $\mathrm{Sn_3N_4}$ to $\mathrm{Si_3N_4}$, where the pressure has chemical origin.

\begin{figure}[t]
\includegraphics[scale=0.35]{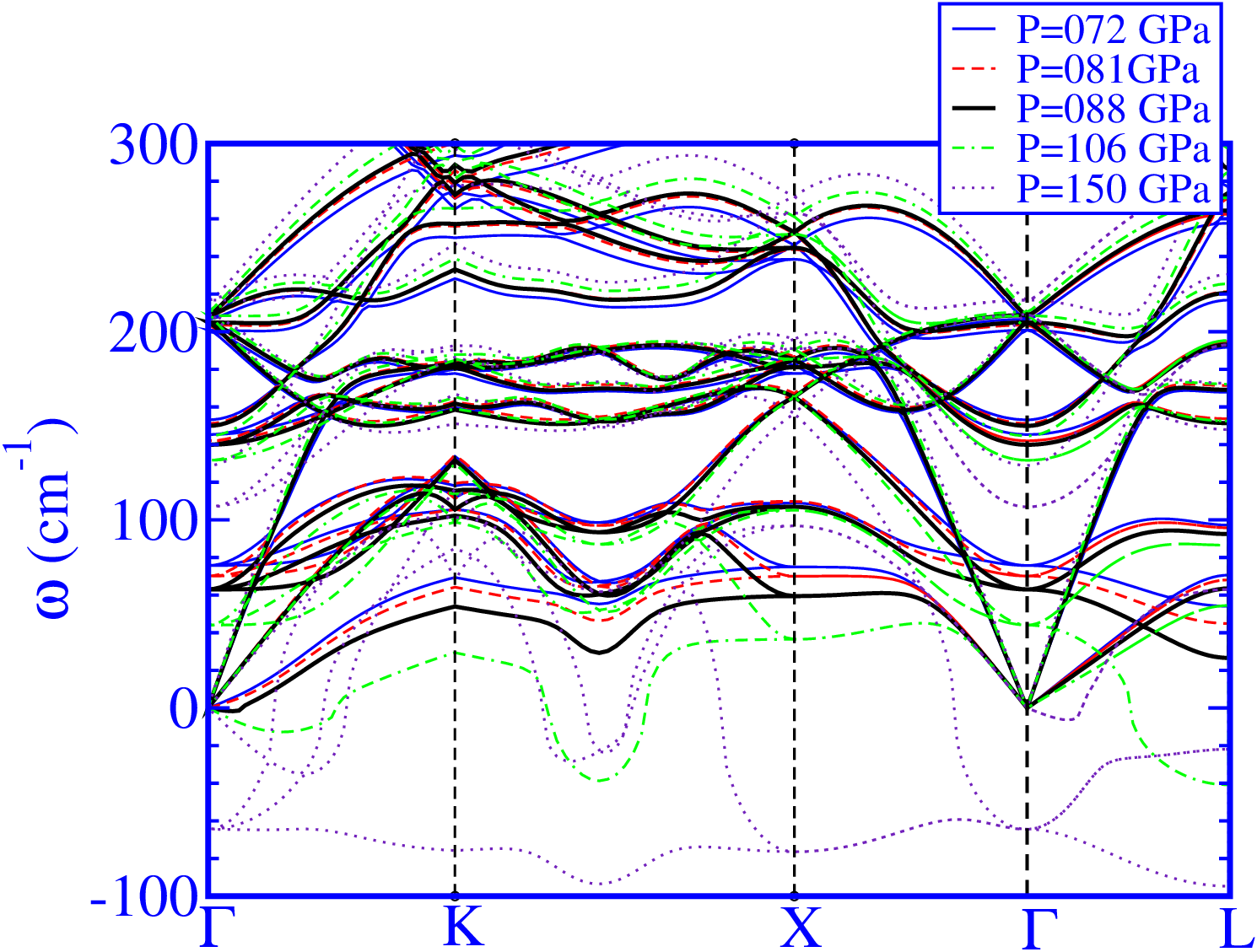}
\caption{Phonon dispersion of $\gamma$-$\mathrm{Sn_3N_4}$ as a function of pressure}
\end{figure}

In summary, in-situ angle dispersive high pressure x-ray studies reveal that bulk modulus of $\gamma$-$\mathrm{Sn_3N_4}$ is 145 ($\pm$1.7) GPa and is stable up to a pressure of 26 GPa. Our first-principles calculations are in excellent agreement with these results with an estimate of a bulk modulus of 158 GPa, well below the earlier calculations. Thus, we demonstrate a significant increase in the compressibility for $\gamma$-$\mathrm{Sn_3N_4}$, compared to lighter members (Si and Ge) in the family. We predict a structural phase transition in $\gamma$-$\mathrm{Sn_3N_4}$ at 88 GPa based on phonon dispersion calculation as a function of pressure. While our calculations confirm a very high value of bulk modulus of $\mathrm{C_3N_4}$ (as predicted earlier)\cite{carbon}, our finding of unstable phonon modes and anomalous Born effective charge show that $\gamma$-$\mathrm{C_3N_4}$ is unstable at ambient conditions. This is in agreement with no experimental reports so far on $\mathrm{C_3N_4}$ synthesized in the cubic form.

GKP, SKD and CN thank ICTP for the financial support under the ICTP-Elettra users programme to carry out this experiment at Elettra, Trieste, Italy. GKP thanks Dr. Diptikanta Swain for useful discussions. SKD thanks Dr. M. P. Shemkunas for providing the samples.


\begin{thebibliography}{10}
\bibitem{tin} N. Scotti, W. Kockelmann, J. Senker, S. Trassel, and H. Jacobs, Z. Anorg. Allg. Chem.
     \textbf{625}, 1435 (1999).

\bibitem{ger} G. Serghiou, G. Miehe, O. Tschauner, A. Zerr, and R. Boehler, J. Chem. Phys.
    \textbf{111}, 4659 (1999).

\bibitem{sil} A. Zerr, G. Miehe, G. Serghiou, M. Schwarz, E. Kroke, R. Riedel, H. Fue$\beta$, P. Kroll and
    R. Boehler, Nature \textbf{400}, 340 (1999).

\bibitem{cohen} A. Y. Liu and M. L. Cohen, Science \textbf{245}, 841 (1989).

\bibitem{carbon} S. D. Mo, L. Ouyang, W. Y. Ching, I. Tanaka, Y. Koyama and R. Riedel, Phys. Rev. Lett.
    \textbf{83}, 5046 (1999)

\bibitem{nano} M. P. Shemkunas, W. T. Petuskey, A. V. G. Chizmeshya, K. Leinenweber, and G. H.
    Wolf, J.  Mat. Res. \textbf{19}, 1392 (2004).

\bibitem{tnsem} Y. Inoue, M. Nomiya and O. Takai, Vacuum \textbf{51}, 673 (1998).

\bibitem{tnel}  O.Takai, Proc. S. I. D. \textbf{28}, 243 (1987).

\bibitem{tnor} T. Maruyama and T. Morishita, Appl. Phys. Lett. \textbf{69}, 890 (1996).

\bibitem{si1b} J. Z. Jiang, H. Lindelov, L. Gerward, K. Stahl, J. M. Recio, P. Mori- Sanchez, S. Carlson,
      M. Mezouar, E. Dooryhee, A. Fitch and D. J. Frost, Phys. Rev. B \textbf{65}, 161202 (2002).

\bibitem{si2b} E. Soignard, M. Somayazulu, J. Dong, O. F. Sankey and P. F. McMillan, J. Phys. Cond.
      Mat. \textbf{13}, 557 (2001).

\bibitem{geb} K. Leinenweber, M. O'Keeffe, M. Somayazulu, H. Hubert, P. F. McMillan and G. H.
     Wolf, Chem. - Eur. J. \textbf{5}, 3076 (1999).

\bibitem{tn1} M. Huang and Y. P. Feng, J. Appl. Phy. \textbf{96}, 4015 (2004).

\bibitem{tn2} W. Y. Ching, S. D. Mo, I. Tanaka and M. Yoshiya, Phys. Rev. B \textbf{63}, 641021 (2001).

\bibitem{tn3} W. Y. Ching, S. D. Mo, L. Z. Ouyang, P. Rulis, I. Tanaka and M. Yoshiya, J. Am. Cer.
      Soc. \textbf{85}, 75 (2002).

\bibitem{tn4} W. Y. Ching and P. Rulis, Phys. Rev. B \textbf{73}, (2006).

\bibitem{tnsynt} M. P. Shemkunas, G. H. Wolf, K. Leinenweber and W. T. Petuskey, J. Am. Cer. Soc. \textbf{85}, 101 (2002).

\bibitem{FIT2D} A. P. Hammersley, S. O. Svensson, M. Hanfland, A. N. Fitch and D. Hausermann,
      High Press. Res. \textbf{14}, 235 (1996).

\bibitem{gold} A. Dewaele, P. Loubeyre and M. Mezouar, Phys. Rev. B \textbf{70}, 094112 (2004).

\bibitem{gga} D. C. Langreth and J. P. Perdew, Phys. Rev. B \textbf{21}, 5469 (1980).

\bibitem{PW} S. Baroni \emph{et al}., \url{http://www.pwscf.org} (2001).

\bibitem{soft} D. Vanderbilt, Phys. Rev. B \textbf{41}, 7892 (1990).

\bibitem{mesh} H. J. Monkhorst and J. D. Pack, Phys. Rev. B \textbf{13}, 5188 (1976); ibid \textbf{16}, 1748 (1977).

\bibitem{GSAS} A. C. Larson and R. B. Von Dreele, GSAS: General Structure Analysis System, Los
     Alamos National Laboratory, Los Alamos, NM (1998).

\bibitem{BM} F. Birch, Phys. Rev. \textbf{71}, 809 (1947).

\bibitem{umesh} U. V. Waghmare, N. A. Hill, H. Kandpal and R. Seshadri, Phys. Rev. B,  \textbf{67}, 125111
      (2003).

\bibitem{ferro} K. M. Rabe and U. V. Waghmare, Phys. Rev. B \textbf{52}, 13236 (1995).

\end{thebibliography}
\end{document}